\documentclass[%
prl,
aps,
twocolumn, 
10pt,
amsmath,amssymb,
noeprint
]{revtex4-2}

\usepackage{xspace}
\usepackage{graphicx}
\usepackage[dvipsnames]{xcolor}
\definecolor{tabblue}{HTML}{1F77B4}
\definecolor{tabpurple}{HTML}{9467BD}
\definecolor{tabpink}{HTML}{e377C2}
\definecolor{tabgreen}{HTML}{2CA02C}
\definecolor{tabgray}{HTML}{7F7F7F}
\usepackage{dcolumn}
\usepackage{bm}
\usepackage{hyperref}
\hypersetup{
    colorlinks=true,
    linkcolor=tabblue!80!black,
    filecolor=tabblue!80!black,      
    urlcolor=tabblue!80!black,
    citecolor=tabblue!80!black
}

\usepackage{orcidlink}
\usepackage{tikz}
\usepackage{etoolbox}
\usepackage{derivative}
\usepackage{verbatim}

\let\oldesp\&                                             
\renewcommand{\&}{{\fontfamily{ppl}\selectfont\oldesp}}

\begin{document}

\title{
    Analytical phase boundary of a quantum driven-dissipative\texorpdfstring{\\}{}Kerr oscillator from classical stochastic instantons
} 

\author{Th{\'e}o S{\'e}pulcre\,\orcidlink{0000-0002-2434-4487}}
\email{theo.sepulcre@riken.jp}
\altaffiliation[\textit{present address:} ]{Center for Quantum Computing, RIKEN, Wako-shi, Saitama 351-0198, Japan.}
\affiliation{Department of Microtechnology and Nanoscience, Chalmers University of Technology, 412 96 Gothenburg, Sweden}

\date{\today}

\begin{abstract}
The framework of Keldysh path integral concisely describes quantum systems driven away from thermal equilibrium, such as the two-photon driven Kerr oscillator. 
Within the thermodynamic limit of diverging photon number, we map it to a Martin-Siggia-Rose-Janssen-de Dominicis path integral, and obtain a purely classical, stochastic equivalent where photon self-interaction plays the role of temperature. 
This perspective sheds light on the difficulties encountered in the search for an effective thermodynamic potential to describe the bistability of the model. 
It allows us to estimate the bistable tunneling rates using a real-time instanton technique leading to an analytical expression of the phase boundary, the first to our knowledge. 
It opens the way to powerful semi-analytical techniques to be applied to various quantum optics models displaying bistability.
\end{abstract}
\maketitle

The familiar experience of water suddenly boiling into vapour has been a long standing puzzle of nineteenth century physics. 
A quantitative understanding came with a pioneer theory of particles in interaction: the Van der Waals equation of state. 
On this basis, Maxwell formulated his famous construction of the liquid-gas phase boundary~\cite{clerk-maxwell_1875_DynamicalEvidence}. 
What seemed to be a peculiarity soon revealed itself in many instances where a macroscopic number of degrees of freedom are collectively interacting. 
Magnets, superconductors or liquid crystals, their variety was hiding a universal behavior, such that one could have borrowed Poincaré's words~\cite{poincare_1908_ScienceMethode} and declare ``Condensed matter physics is the art of giving the same name to different things''. 
Thermal fluctuations can be replaced by quantum ones~\cite{sachdev_2011_QuantumPhase}, to create disordered phases of matter by virtue of tunneling events between degenerate ground states, restoring symmetries that classical dynamics had to break. 
In that case, the quantum vacuum replaces thermal equilibrium. 
Even further, recent work has shown that driven and dissipative quantum systems—operating far from equilibrium—can exhibit analogous transitions~\cite{dallatorre_2010_QuantumCritical,sieberer_2013_DynamicalCritical,sieberer_2016_KeldyshField}. 
Whether such driven–dissipative phenomena can be placed within the same universality classes as their thermal counterparts, and how to characterize their phase transitions, remain open and compelling questions.

Perhaps the simplest example of such a quantum driven-dissipative model is a single photonic mode subject to photon-photon interaction \(U\), two-photon drive \(\epsilon\) and losses \(\gamma\), summarized in the Hamiltonian 
\begin{equation}\label{eq:hamiltonian}
    \hat{H} = 
    \delta \hat{a}^{\dagger} \hat{a} 
    + \frac{U}{2} \hat{a}^{\dagger 2} \hat{a}^{2} 
    + \frac{\epsilon}{2} \left(\hat{a}^{\dagger 2} + \hat{a}^{2}\right).
\end{equation}
Dissipation is described by the Lindblad jump operator \(\hat{L} = \sqrt{2 \gamma} \hat{a}\)~\cite{breuer_2002_TheoryOpen}. 
We work here in the frequency frame of the drive, such that \(\delta=\omega_0 - \omega_{d}\), \(\omega_0\) the mode bare frequency and \(\omega_{\mathrm{d}}\) the drive frequency. 

This model or its single-photon driven variant has been extensively studied in quantum optics: it is well known to display bistability~\cite{drummond_1980_QuantumTheory}, \textit{i.e.}, the coexistence of two stable steady-state solutions at mean-field level, which seemed to contradict the necessary uniqueness of the quantum ground state. 
At vanishing interaction, the photon number diverges, signaling the approach to a thermodynamic limit, and leading to a genuine first-order phase transition across the bistable regime, with exponentially high tunneling rates between metastable states\cite{beaulieu_2025_ObservationFirst,castillo-moreno_2025_ExperimentalObservation}. 
The resultant hysteresis has been observed~\cite{casteels_2016_PowerLaws,rodriguez_2017_ProbingDissipative}. 
It has been proposed as a basis for critical sensing~\cite{petrovnin_2024_MicrowavePhoton,beaulieu_2025_ObservationFirst}. 
Similar behavior is also reported for resonator arrays numerically~\cite{vicentini_2018_CriticalSlowing} and experimentally~\cite{muppalla_2018_BistabilityMesoscopic,andersen_2020_QuantumClassical}.
Yet, despite decades of studies, including the dicovery of analytical solutions for the photon number at finite \(U\)~\cite{kryuchkyan_1996_ExactQuantum,kheruntsyan_1999_WignerFunction,minganti_2016_ExactResults,roberts_2020_DrivendissipativeQuantum}, a satisfactory formulation of a thermodynamic potential---which could locate the phase boundary using Maxwell's construction---has proven hard to find~\cite{muppalla_2018_BistabilityMesoscopic,andersen_2020_QuantumClassical,zhang_2021_DrivendissipativePhase,peters_2023_ScalarPotentials,mylnikov_2025_EmergentEquilibrium,petrovnin_2025_NumericalSimulation}.

\paragraph{Semi-classical limit.} The framework of Keldysh path integral has been recently appraised to tackle such quantum driven-dissipative problems~\cite{maghrebi_2016_NonequilibriumManybody,sieberer_2016_KeldyshField,zhang_2021_DrivendissipativePhase,thompson_2023_FieldTheory}.
It especially provides non-perturbative methods to compute bistable tunneling times~\cite{titov_2016_KorshunovInstantons,thompson_2022_QubitDecoherence,lee_2025_RealtimeInstanton,carde_2025_NonperturbativeSwitching}
The partition function \(Z = \int \mathfrak{D}[\alpha_{\mathrm{c}}, \alpha_{\mathrm{q}}, \alpha^*_{\mathrm{c}}, \alpha^*_{\mathrm{q}}] e^{i\mathcal{S}[\alpha_{\mathrm{c}}, \alpha_{\mathrm{q}}]}\) is the key quantity from which any observable is derived. 
The field \(\alpha_{\mathrm{c}}\) represents the average values accumulated by operators, while \(\alpha_{\mathrm{q}}\) encodes the quantum fluctuations. 
The Keldysh action corresponding to Eq.~\eqref{eq:hamiltonian} for an evolution time \(\tau\) is
\begin{align}
    &\mathcal{S} = \int_0^\tau{\rm d}t\, 
    \left( 
        \alpha^*_{\mathrm{c}}(i \partial_t - \delta - i\gamma)\alpha_{\mathrm{q}} + {\text{\footnotesize H.c.}}
    \right) + 2i\gamma |\alpha_{\mathrm{q}}|^2\nonumber \\
	&- \frac{U}{2}\left(
        |\alpha_{\mathrm{c}}|^2 + |\alpha_{\mathrm{q}}|^2
        \right)(\alpha_{\mathrm{c}}^*\alpha_{\mathrm{q}} + {\text{\footnotesize H.c.}})
	- \epsilon(\alpha^*_{\mathrm{c}}\alpha^*_{\mathrm{q}} + {\text{\footnotesize H.c.}}),
\end{align}
We perform a semi-classical scaling of the integration field variables, \(\alpha_{\mathrm{c}} \rightarrow  \alpha_{\mathrm{c}} \sqrt{{\gamma}/{U}}\) and \(\alpha_{\mathrm{q}} \rightarrow \alpha_{\mathrm{q}} \sqrt{{U}/{\gamma}}\). It leaves the integration measure unchanged. 
The quadratic term in \(\alpha_{\mathrm{q}}\) now reads \(2iU |\alpha_{\mathrm{q}}|^2\). In the \(U / \gamma \to 0\) limit, we can neglect the interaction term \(\propto \alpha_{\mathrm{q}}^3\), leaving (\({\gamma}/{2})|\alpha_{\mathrm{c}}|^2(
        \alpha_{\mathrm{c}}^*\alpha_{\mathrm{q}} + {\text{\footnotesize H.c.}}
    )\).
It reduces the Keldysh path integral to a Martin-Siggia-Rose-Janssen-de Dominicis (MSRJD) path integral~\cite{kamenev_2011_FieldTheory,tauber_2007_FieldtheoryApproaches,hertz_2016_PathIntegral,zakine_2023_MinimumactionMethod,altland_2010_CondensedMatter}.
To make it more apparent, we relabel the fields as 2-vectors, \(\vec{q}=\sqrt{2}({\rm Re}\,\alpha_{\mathrm{c}}, {\rm Im}\,\alpha_{\mathrm{c}})\), 
\(\vec{p}= \sqrt{2}U(-i {\rm Im}\,\alpha_{\mathrm{q}}, i {\rm Re}\,\alpha_{\mathrm{q}})/U\).
We finally define a real-valued action \(S\) by \(i\mathcal{S}=-{S}/{U}\). 
It assumes the form
\begin{equation}
    S = \int_0^\tau {\rm d}t\, \vec{p} \cdot \dot{\vec{q}} - H(\vec{q}, \vec{p}),
    \quad 
    H = |\vec{p}|^2 + \vec{p} \cdot \vec{f}(\vec{q}).
\end{equation}
This action describes a classical particle of position \(\vec{q}\) subject to a fluctuation force encoded in the pseudo-momentum \(\vec{p}\) and a deterministic force, 
\begin{equation}\label{eq:force}
    \vec{f}(\vec{q}) = - \left(
        \gamma 
        +(\delta + \gamma|\vec{q}/2|^2) \mathbf{J}
        + \epsilon \mathbf{M}
    \right) \vec{q},
\end{equation}
where \(\mathbf{J}=\left(\begin{smallmatrix}0 & -1\\ 1 & 0\end{smallmatrix}\right)\) and \(\mathbf{M}=\left(\begin{smallmatrix}0 & 1\\ 1 & 0\end{smallmatrix}\right)\). Solving \(\vec{f}(\vec{q})=0\) reproduces mean-field equations.
Since the action is quadratic in \(\vec{p}\), it can be exactly integrated, to obtain a Langevin first-order differential equation for \(\vec{q}\)~\cite{sieberer_2016_KeldyshField}. 
We now compute the probability that this Langevin particle tunnels between two positions \(\vec{q}_{\mathrm{i}}\) and \(\vec{q}_{\mathrm{f}}\). 
Along the paths contributing the most to \(Z\), \(S[\vec{p}, \vec{q}]/U\) is minimal. 
This intuition is made tangible by a saddle point approximation around the least action path, \(q_{\ell}(t),\, p_{\ell}(t)\), which verifies 
\begin{equation} \label{minimalAction}
    \fdv{S}{q}_{p_{\ell}, q_{\ell}} = 
    \fdv{S}{p}_{p_{\ell}, q_{\ell}} = 0.
\end{equation}
The tunnel probability is approximated by \(p(q_{\mathrm{f}}, \tau| q_{\mathrm{i}}, 0) \propto \exp(-S[\vec{q}_{\ell}, \vec{p}_{\ell}]/U)\). 
In this Boltzmann-like activation factor, the interaction energy \(U\) plays the role of an effective temperature. 
The minimal action condition is nothing but Hamilton's equation of motion, \(\dot{\vec{q}} = \partial_{\vec{p}}H = 2\vec{p} + \vec{f}(\vec{q})\) and \( \dot{\vec{p}} = -\partial_{\vec{q}}H\).
To compute the tunneling rate \(p\), one has to solve these differential equations and evaluate the action along these paths, called instantons, by analogy with quantum tunneling.

The scenario for this first-order transition is now clear. 
When \(U / \gamma \to 0\), this quantum driven-dissipative model reduces to a classical stochastic system~\cite{foss-feig_2017_EmergentEquilibrium}, where the interaction strength \(U\) weights the quantum fluctuations.  
In the bistable regime, they trigger tunneling events between a vacuum and a bright metastable state. 
At vanishing fluctuations, \textit{i.e.}, at the thermodynamic limit~\footnote{The photon occupation number also diverges when \(U / \gamma \to 0\), which can be interpreted as the large number of degrees of freedom usually expected at the thermodynamic limit}, the tunneling events become exponentially rare. 
It leads to hysteresis, and finally to a discontinuity of \(\langle\hat{a}\rangle\) when the tunneling rates are equal going from vacuum to bright state and back, or equivalently when the minimal action is equal in both directions. 

One could recast this MSRJD path integral into a Fokker-Plank equation, and connect with existing methods~\cite{bonifacio_1978_PhotonStatistics,vogel_1989_QuasiprobabilityDistributions,kryuchkyan_1996_ExactQuantum,kheruntsyan_1999_WignerFunction}.
The corresponding Langevin process can also be sampled numerically, leading to the so-called Truncated Wigner Approximation~\cite{vogel_1989_QuasiprobabilityDistributions,carusotto_2005_SpontaneousMicrocavitypolariton,carusotto_2013_QuantumFluids,singh_2022_DrivendissipativeCriticality,mink_2022_VariationalTruncated}.
This approach benefits from recent advances on stochastic differential solvers~\cite{rackauckas_2017_AdaptiveMethods}. 

\paragraph{Ornstein-Uhlenbeck pseudo-potential.}

\begin{figure}
    \centering
    \includegraphics{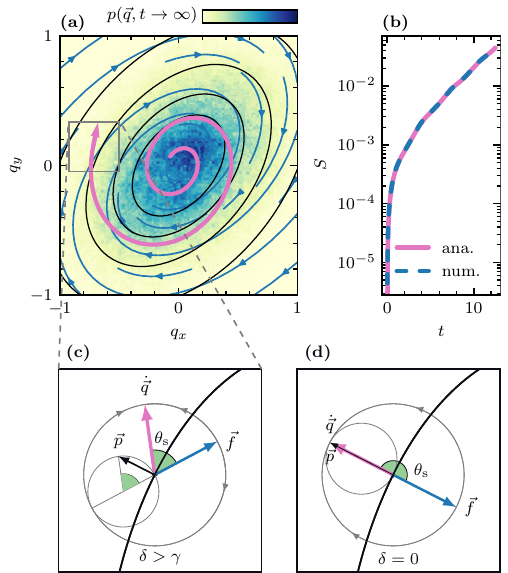}
    \caption{\textbf{Minimal action paths of the Ornstein-Uhlenbeck model.} \textbf{(a)} Escape trajectory obtained by numerical integration (pink), non-conservative force field (blue) and equipotentials of \(P\) (black).
    The steady-state is obtained by solving and averaging the Langevin dynamics (colormap).
    \textbf{(b)} Action along the instanton, using analytical expression based on Eq.~\eqref{eq:OUpseudopotential} (pink) and numerical integration (blue).
    \textbf{(c)} Given \(\vec{f}\), \(\vec{p}\) and \(\dot{\vec{q}}\) are constrained to the grey circles.
    Grey arrows indicate the dynamical flow of \(\theta(t)\), see Eq.~\eqref{eq:theta}.
    \textbf{(d)} In a pure potential force (\(\delta=0\)), \(\theta_{\rm s}=\pi\), and the escape trajectory opposes \(\vec{f}\).}
    \label{fig1}
\end{figure}

We start building physical intuition by studing simpler cases. 
One dimensional systems are simple enough to be integrated exactly. 
Firstly, \(H\) is conserved by Hamilton's equations; in the \(\tau\to\infty\) limit, the minimal action path must obey \(H=0\). 
Two types of trajectories are then possible. 
Either \(p=0\) such that \(\dot{q}=f\); it generates a zero-action trapping trajectory towards a metastable state. Or \(p=-f\,\Rightarrow\,\dot{q} = -f\), an escape trajectory driven by stochastic fluctuations. 
Secondly, the force always derives from a potential, \(f(q) = -U'(q)\), such that the action does not depend on the integration path, \(S=U(q_{\mathrm{f}}) - U(q_{\mathrm{i}})\). 
This is the typical framework of Kramer's theory~\cite{hanggi_1990_ReactionrateTheory,muppalla_2018_BistabilityMesoscopic}. 

In two dimensions, the situation is more intricate, since the \(H=0\) condition does not entirely constrain \(\vec{p}\).
We parametrize the constraint with a new dynamic parameter \(\theta\), such that 
\begin{equation}\label{eq:pconstraint}
    H=0 \quad\Leftrightarrow\quad\vec{p} = \frac{1}{2}\left( e^{\theta \mathbf{J}} - 1 \right) \vec{f}.
\end{equation}
As a consequence, \(\dot{\vec{q}}= e^{\theta \mathbf{J}} \vec{f}\). 
Secondly, the force will not in general derive from a potential. 
At best, we can parametrize it with a pure gradient part \(U(\vec{q})\) and a pure curl \(V(\vec{q})\) as 
\begin{equation}
    \vec{f}(\vec{q}) = -\vec{\nabla}U(\vec{q}) - \mathbf{J}\vec{\nabla}V(\vec{q}).
\end{equation}
Under this decomposition, the differential equation for \(\theta(t)\) becomes 
\begin{equation}\label{eq:theta}
    \dot{\theta}=\sin(\theta) {\Delta}U(\vec{q}) - (1-\cos(\theta)){\Delta}V(\vec{q}).
\end{equation}
By analyzing the stationary solutions \(\dot{\theta}=0\), we obtain the minimal action path classification for 2D dynamics. 
The first stationary solution \(\theta_{\mathrm{s}} = 0\), generates a trapping trajectory, \(\dot{\vec{q}} = \vec{f}\). 
The second stationary solution, at which the particle follows an escape path, is found at \(\theta_{\rm s} = 2 \arctan(\Delta U / \Delta V)\). 
Unless the force derives from a gradient (\(V=0\)), this stationary angle now depends on \(\vec{q}\). [Fig.~\ref{fig1}(c)\,\&\,(d)].
This classification can be illustrated by the Ornstein-Ulhenbeck (OU) process~\cite{uhlenbeck_1930_TheoryBrownian}, the stochastic process obtained if \(\vec{f}\) is at most linear in \(\vec{q}\), \textit{e.g.}, by suppressing the interaction term in Eq.~\eqref{eq:force}. 
The force writes 
\begin{equation}
    \vec{f}(\vec{q}) = -(\gamma + \delta \mathbf{J} + \epsilon \mathbf{M}) \vec{q}.
\end{equation}
This expression parametrizes any \(2 \times 2\) matrix in terms of our model parameter, up to a rotation of the reference frame of \(\vec{q}\). 
Here, \(\delta\) weights the curl part of the force: it does not derive from a gradient. 
The stationary angle generating the escape trajectory is \(\tan\theta_{\mathrm{s}} /2 = \gamma / \delta\). 
We now assume that the dynamics of \(\theta(t)\) is fast compared to \(\vec{q}(t)\), such that \(\theta(t)=\theta_{\mathrm{s}}\). 
We can fully express \(\vec{p}\) as a function of \(\vec{q}\). It now derives from a pseudo-potential, \(\vec{p} = \vec{\nabla}P(\vec{q})\) with 
\begin{equation}\label{eq:OUpseudopotential}
    P(\vec{q}) = \frac{\cos \theta_\nu}{2}
    \vec{q} \cdot \left( 
        \nu + \epsilon \mathbf{M} e^{\theta_\nu \mathbf{J}} 
    \right) \vec{q},
\end{equation}
where we used polar form \(\gamma=\nu \cos \theta_{\nu}\), \(\delta = \nu \sin \theta_\nu\). 
The action is now independent of the path, \(S = P(\vec{q}_{\mathrm{f}}) - P(\vec{q}_{\mathrm{i}})\). 
Consequently, we obtain a Boltzmann factor-like steady-state density probability \(p(\vec{q}) \propto \exp (-P(\vec{q}) / U)\). 
The situation is illustrated on Fig.~\ref{fig1}, where we drew the equipotential lines of \(P\), the non-conservative force field, and an escape minimal action path in Fig.~\ref{fig1}\,(a). 
We independently obtained the steady-state by simulation of the corresponding Langevin dynamics, using an Euler-Maruyama integration scheme. 
We furthermore confirm the validity of \(\theta(t) \simeq \theta_{\mathrm{s}}\) by comparing the action along the path computed with the pseudo-potential \(P\) and by numerical integration of the equation of motion.

\paragraph{Approximation schemes.}
\begin{figure}
    \centering
    \includegraphics{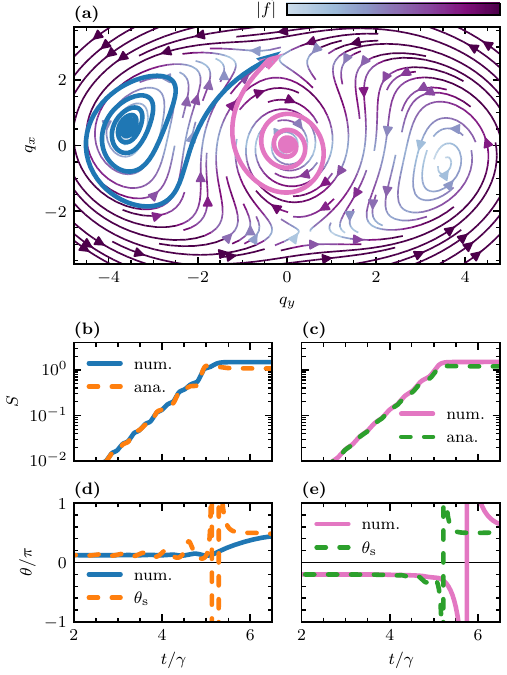}
    \caption{\textbf{Minimal action paths of the 2-photon driven Kerr oscillator model} at \(\delta / \gamma = -10\) and \(\epsilon / \gamma = 3.2\). 
    \textbf{(a)} Instanton escape trajectories from the vacuum state (pink) and the bright state (blue). 
    The streamlines represent the non-conservative force field \(\vec{f}\). 
    \textbf{(b)} Action along the bright state escape path, by numerical integration (blue) and analytical ansatz (orange). 
    \textbf{(c)} Action along the vacuum state escape path, by numerical integration (pink) and analytical ansatz (green). 
    \textbf{(d)} \(\theta(t)\) along the bright state escape, by numerical integration (blue) and value of the instantaneous stationary angle \(\theta_{\rm s}(\vec{q})\) (orange). 
    \textbf{(e)} \(\theta(t)\) along the vacuum state escape, by numerical integration (pink) and value of the instantaneous stationary angle \(\theta_{\rm s}(\vec{q})\) (green).}
    \label{fig2}
\end{figure}

Aside from special cases like the OU process, there is no guarantee that \(\vec{p}\) derives from a gradient when \(\theta(t) = \theta_{\mathrm{s}}\). It is especially not the case concerning the driven-dissipative Kerr oscillator, where the force field is given by Eq.~\eqref{eq:force}.
We now understand the inherent difficulties of the model at hand: the non-linear term is a pure curl, with no associated potential, such that supplementary approximations are required to analytically compute tunneling rates. 
To gain intuition, we employ numerical integration to find minimal action escape paths connecting stable fixed points to unstable fixed point of \(\vec{f}\) [Fig.~\ref{fig2}\,(a)] \footnote{The full tunneling trajectory will continue \textit{via} a trapping, actionless path towards the next stable fixed point, and is ignored in our analysis.}. 
Standard approaches to this boundary value problem like the shooting method~\cite{press_2007_NumericalRecipes} are unreliable, because the strong oscillatory trajectory is extremely sensitive to initial conditions in the \(\tau\to\infty\) limit. 
We refine it by raising the total instanton time \(\tau\) incrementally. 
At every step, we solve the shooting problem, using the previous step solution as an initial guess. 
This allows the algorithm to progressively adapt to the details of the force field and to the trajectory fast oscillations. More advanced methods have been explored in the literature (see \cite{zakine_2023_MinimumactionMethod} and references therein).  
We observe that the condition \(\theta(t) = \theta_{\mathrm{s}}\) is obeyed in the vicinity of the fixed points, with significant deviations when the trajectory crosses from one attraction basin to another [Fig.~\ref{fig2}\,(d)\,\&\,(e)].
We thus formulate an approximation amenable to analytical computations, assuming that \(\theta(t) = \theta(-\infty)\), \textit{i.e.}, \(\theta\) stays locked at the value necessary to escape its initial trap.
Eq.~\eqref{eq:pconstraint} now completely determines \(\vec{p}(\vec{q})\).
We discard its curl part, and write it as a gradient.
It should be noted that this pseudo-potential, while being a technical convenience, now depends on the initial state.
The action computed numerically and analytically within these approximations are compared on Fig.~\ref{fig2}\,(d)\,\&\,(e).
As expected, they match during the initial escape, but discrepancies accumulate towards the end of the trajectory. 

\begin{figure}
    \centering
    \includegraphics{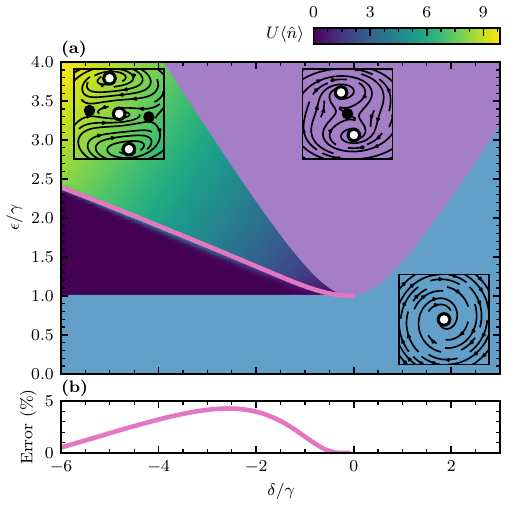}
    \caption{\textbf{2-photon driven Kerr oscillator phase diagram.} \textbf{(a)} The mean-field phase diagram in detuning \(\delta\) and drive amplitude \(\epsilon\) shows a vacuum regime (blue), a cat regime (purple) and a bistable regime (colormap).
    Insets show stable (white) and unstable (black) states, and \(\vec{f}\) aspect in each region.
    The colormap shows the photon number in the bistable regime at \(U = 0.1 \gamma\) using exact expressions.
    A phase boundary appears, with an abrupt change of the photon number, separating a vacuum and a bright phase.
    The analytical phase boundary obtained at Eq.~\eqref{eq:transitionLine} (pink) matches.
    \textbf{(b)} The relative difference between transition boundary predicted by instanton method and numerically located on exact expression for photon number.}
    \label{fig3}
\end{figure}

Finally, we test this analytical approach by computing the phase boundary, as the \textit{locus} of \((\delta / \gamma, \epsilon / \gamma)\) points where the action (and thus the escape probability) from the outer and inner fixed points (corresponding to bright and vacuum states respectively) are equal. 
We find it defined by the implicit equation 
\begin{align}\label{eq:transitionLine}
    &\frac{
        \delta (\sqrt{\epsilon^2 - \gamma^2} + \delta)^2
    }{
        4 \sqrt{\epsilon^2-\gamma^2} (\gamma^2+ \delta^2)
    }(\gamma^2+ (2 \sqrt{\epsilon^2 - \gamma^2} - \delta)^2)  \nonumber\\
    =&\epsilon^2 - \gamma^2 - \left(
        \delta - \sqrt{\epsilon^2 - \gamma^2} 
        + \gamma\sqrt{\frac{\delta^2 - \epsilon^2 + \gamma^2 }{\epsilon^2}}
    \right)^2.
\end{align}
This line is shown on Fig.~\ref{fig3} on top of the numerically computed phase diagram of the system within the bistable region.
The relative error between the analytical line and numerical results remains below 5\% throughout.
It vanishes when \(\delta / \gamma \to -\infty\), when the particle spends the longest time escaping its initial metastable point, as well as at \(\delta / \gamma \to 0\), since the curl part of the force field is negligible.

The onset of this first order phase transition has been observed \textit{e.g.} on circuit QED based platforms, \textit{via} heterodyne detection of the emitted field, either in average~\mbox{\cite{petrovnin_2024_MicrowavePhoton}}, or resolved as a time series of tunneling events~\mbox{\cite{beaulieu_2025_CriticalityenhancedQuantum}}. 
Emitted power spectrum measurement~\mbox{\cite{castillo-moreno_2025_ExperimentalObservation}} or dynamical hysteresis area law~\mbox{\cite{rodriguez_2017_ProbingDissipative,casteels_2016_PowerLaws}} are also promissing avenues. 
To our knowledge, no comparison with a theoretically obtained boundary line as been attempted so far on this model. 
The precise measure of the line position is delicate: close to the critical point, approaching the thermodynamic limit to obtain a sharp transition requires vanishingly small $U/\gamma$ ($\sim 10^{-4}$ reported by \mbox{\cite{petrovnin_2024_MicrowavePhoton}}). 
Thermal noise is then competing with quantum fluctuations to activate the tunneling events and broadens the transition line. 
Conversely, at high power, the exponentially large times between switches dictate unpractically long integration times.

\paragraph{Conclusion.} 
We have derived a classical stochastic description of the two-photon-driven Kerr oscillator in the thermodynamic limit by mapping the Keldysh path integral to its MSRJD counterpart. 
This mapping reveals how quantum fluctuations, governed by the interaction strength \(U\), act as an effective temperature.

Using a real-time instanton approach, we obtained a novel analytical expression for the first-order transition line between bright and dim phases. 
Our results provide a clear physical picture of bistability in this system and explain the failure of several previous pseudo-potential approaches. 
Furthermore, this framework unifies several established methods used for driven-dissipative systems---including the Truncated Wigner and Fokker-Planck techniques---and justifies the approximations on which they rely.

The validity of the approach must be examined further around the critical points at \(\delta>0\), where divergingly large fluctuations usually defeat mean-field methods. 
Here, they challenge the neglect of the cubic terms in the quantum fields. 
A careful simulation of the dynamical critical exponent would offer a useful benchmark of deviation from mean-field predictions. 

Finally, we acknowledge that the single Kerr oscillator is a simple testbed, amenable to brute-force diagonalisation. 
Yet, we believe that the method exposed here will scale favorably to more complex systems, where similar first-order driven-dissipative transitions are expected, \textit{e.g.}, driven Bose-Hubbard arrays or driven Tavis-Cummings model from quantum optics~\cite{bonifacio_1978_PhotonStatistics}.
The separation between fast \(\theta(t)\) and slow \(\vec{q}(t)\) variables will prove essential for addressing truly many-body systems by reducing the number of degrees of freedom.
Together with numerical techniques imported from classical stochastic processes~\cite{zakine_2023_MinimumactionMethod}, it significantly broadens the toolset available to study quantum driven-dissipative phase transitions.

\paragraph{Acknowledgements.}
The author thanks Vitaly Shumeiko for bringing the problem of the phase boundary to his attention, as well as Therese Karmstrand and Timo Hillmann for fruitful discussions. 
This work received support from the Knut and Alice Wallenberg Fundation through the Wallenberg Center for Quantum Technology (WACQT).

\bibliography{MSR}

\end{document}